%% file: ms.tex
\documentclass[letterpaper,twocolumn,10pt]{article}
\usepackage{usenix-2020-09}
\usepackage[toc,page]{appendix}

\usepackage{tikz}
\usepackage{amsmath}
\usepackage{listings, listings-rust}
\usepackage[scaled=0.85]{beramono}
\usepackage{xurl}
\begin{document}

\date{}

\title{\Large \bf Maybenot: 
A Framework for Traffic Analysis Defenses}

\author{
{\rm Tobias Pulls} \\
Karlstad University\\
{\rm tobias.pulls@kau.se}
\and
{\rm Ethan Witwer} \\
Link{\"o}ping University\\
{\rm ethan.witwer@liu.se}
}

\maketitle

\begin{abstract}
\input{src/abstract.tex}
\end{abstract}

\section{Introduction}
\label{sec:intro}
\input{src/intro.tex}

\section{Background}
\label{sec:background}
\input{src/background.tex}

\section{Maybenot Framework}
\label{sec:framework}
\input{src/framework.tex}

\section{Maybenot Machines}
\label{sec:machines}
\input{src/machines.tex}

\section{Maybenot Simulator}
\label{sec:simulator}
\input{src/simulator.tex}

\section{Discussion}
\label{sec:discussion}
\input{src/discussion.tex}

\section{Related Work}
\label{sec:related}
\input{src/related.tex}

\section{Conclusion}
\label{sec:conclusion}
\input{src/conclusion.tex}

\section*{Acknowledgments}
For their valuable feedback, we would like to thank Matthias Beckerle, Anna Brunström, Niklas Carlsson, Rasmus Dahlberg, Gr{\'{e}}goire Detrez, Johan Garcia, Nick Hopper, Joakim Hulthe, Linus Färnstrand, David Hasselquist, Jan Jonsson, Oscar Linderholm, David Lönnhager, Nick Mathewson, Daniel Paoliello, Mike Perry, Markus Pettersson, Florentin Rochet, Odd Stranne, Fredrik Strömberg, Björn Töpel.

This work was supported by Mullvad VPN, the Swedish Internet Foundation, the Knowledge Foundation of Sweden, Karlstad Internet Privacy Lab, and the Wallenberg AI, Autonomous Systems and Software Program (WASP) funded by the Knut and Alice Wallenberg Foundation. The paper was written using Github Copilot and Grammarly.

\section*{Availability}
The Maybenot framework and simulator are available at \url{https://crates.io/crates/maybenot} and \url{https://crates.io/crates/maybenot-simulator}. They are both dual-licensed under either the MIT or Apache 2.0 license. An FFI wrapper is available at \url{https://crates.io/crates/maybenot-ffi} with the same license.

For development, we have bundled the framework, simulator, and other crates related to Maybenot into a workspace at \url{https://github.com/maybenot-io/maybenot}.

\bibliographystyle{plain}
\bibliography{ref}

\end{document}

%% file: src/abstract.tex
End-to-end encryption is a powerful tool for protecting Internet users' privacy. With the increasing use of technologies such as Tor, VPNs, and encrypted messaging, it is becoming increasingly difficult for network adversaries to monitor and censor Internet traffic. One remaining avenue for adversaries is traffic analysis, the analysis of patterns in encrypted traffic to infer information about users and their activities. Recent improvements in deep learning have made traffic analysis attacks more effective than ever before.

We present version 2 of Maybenot, a framework for traffic analysis defenses. Maybenot is designed to be easy to use and integrate into existing end-to-end encrypted protocols. It is implemented in Rust as a crate (library), together with an FFI wrapper to enable the usage of Maybenot in other programming languages and a simulator to further the development of defenses. Defenses in Maybenot are expressed as probabilistic finite-state machines that schedule actions to inject padding or block outgoing traffic. Maybenot is an evolution of the Tor Circuit Padding Framework by Perry and Kadianakis, designed to support a wide range of protocols and use cases.

%% file: src/intro.tex
After decades of effort, end-to-end encryption is becoming ubiquitous, with modern protocols such as QUIC~\cite{quic}, HTTP/3~\cite{http3}, TLS 1.3~\cite{tls13}, DoH~\cite{doh}, DoQ~\cite{doq}, MLS~\cite{ietf-mls-protocol-17}, and more encrypted by default. In parallel, technologies that provide unlinkability between end users and their IP addresses under various threat models are being increasingly adopted, such as Tor~\cite{tor}, VPNs~\cite{vpnadopt, vpnadopt2, wg}, and Apple's iCloud Private Relay~\cite{privaterelay,ietf-masque-connect-ip-07}. Together, these two trends are making it harder for network operators to detect and block malicious traffic and for attackers to identify and target specific users. The final frontier is traffic analysis: inferences based on the metadata of encrypted traffic.

While traffic analysis attacks have been studied for a long time~\cite{steppingstones,ap,Danezis04,MurdochD05,KedoganAP02}, we see relatively few deployed real-world defenses. Most defenses can be found in technologies aiming for user--IP address unlinkability, such as Tor~\cite{tor}; however, they remain relatively modest in induced overhead~\cite{circpad}, limiting their efficacy against many attacks. Recent protocol standards do support essential building blocks and note that they may be used to defend against traffic analysis (cf. support for the \texttt{PADDING} frame in QUIC~\cite[§21.14]{quic} and reflections on traffic analysis in TLS 1.3~\cite[§3]{tls13}), yet they do not mandate any specific use.

There are several reasons for the lack of deployed traffic analysis defenses. For one, the negative performance impact of padding and added latency is a significant concern: with high costs, the benefits must be very clear. This is compounded by the fact that the landscape of traffic analysis attacks and defenses is rapidly evolving, driven by advances in deep learning and artificial intelligence~\cite{sok}. Additionally, one must recognize that it has taken decades to achieve widespread adoption of end-to-end encryption. Without robust encryption, data/payload leakage has typically been perceived as a more significant threat than metadata.

This paper presents version 2 of Maybenot, a \emph{framework} for traffic analysis defenses. Maybenot is designed to be easy to integrate into existing protocols and flexible enough to support a wide range of traffic analysis defenses. Version 1 is also documented on arXiv~\cite{maybenotv1} and partially covered by a publication at WPES 2023~\cite{maybenotwpes}. Maybenot has been integrated with WireGuard by Mullvad VPN into their Defense against AI-guided Traffic Analysis (DAITA)\footnote{\url{https://mullvad.net/en/blog/introducing-defense-against-ai-guided-traffic-analysis-daita}} VPN feature.

Section~\ref{sec:background} provides background on traffic analysis with a focus on website fingerprinting defenses. Section~\ref{sec:framework} presents the Maybenot framework, followed by Section~\ref{sec:machines}, which describes \emph{machines} that run in the framework. Section~\ref{sec:simulator} covers the Maybenot simulator for simulating network traces with Maybenot running at the client and server side. Section~\ref{sec:discussion} discusses limitations and the merits of a framework and motivates the changes made in version 2 of Maybenot. Section~\ref{sec:related} explains related work, and Section~\ref{sec:conclusion} concludes the paper.

%% file: src/background.tex
Maybenot is mainly sprung out of work on website fingerprinting defenses, which is closely related to end-to-end flow correlation/confirmation~\cite{detorrent,deepcorr,deepcoffea,The23rdRaccoon,trilemma,trilemma2}. We therefore focus on website fingerprinting in Section~\ref{sec:background:wf} and the broader impact of traffic analysis defenses in Section~\ref{sec:background:impact}.

\subsection{Website Fingerprinting}
\label{sec:background:wf}
In the website fingerprinting (WF) setting, a local, passive attacker observes encrypted network traffic generated by a client visiting a website using some proxy or relay that hides the destination IP address---typically Tor~\cite{tor}, but it could just as well be a VPN~\cite{cheng1998traffic,touching,SunSWRPQ02,Hintz02,LiberatoreL06,HerrmannWF09} or some developing standard like MASQUE~\cite{ietf-masque-connect-ip-07,privaterelay}.
The goal of the attacker is to infer the website the client is visiting (or the \emph{subpage} of a website rather than the typical \emph{frontpage}, in the case of \emph{webpage} fingerprinting).
Section~\ref{sec:background:wf:attacks} gives an overview of WF attacks and Section~\ref{sec:background:wf:defenses} of WF defenses.

\subsubsection{Attacks}
\label{sec:background:wf:attacks}

WF attacks can be grouped based on how they deal with feature engineering: manually or automatically. Both types of attacks make use of the sequence of packets in a network trace as well as their directions, timestamps, and sizes (if applicable: most attacks are tuned for Tor, where traffic is packed into constant-size cells of 514 bytes\footnote{\url{https://spec.torproject.org/tor-spec/preliminaries.html?highlight=514\#msg-len}}).

In gist, the community spent about a decade discovering the most valuable features for WF, culminating in the k-fingerprinting attack by Hayes and Danezis~\cite{kfp} and CUMUL by Panchenko et al.~\cite{cumul}. They both have on the order of 100--200 features, where the top 10--20 features provide the overwhelming majority of utility.

Unfortunately, progress in the area of deep learning from 2016 and onward has made automatic feature engineering practical and superior to manual feature engineering~\cite{abe2016fingerprinting,RimmerPJGJ18,df,tiktok,varcnn,maturesc,rf}. An attacker trains on raw\footnote{Albeit the trace representation used as input differs---such as a sequence of directional timestamps~\cite{tiktok} or per-direction counts of packets within time windows~\cite{rf}---which is a (minimal) form of feature engineering.} network traffic and uses the learned features for classification. This is a significant improvement over manual feature engineering, as it does not require any domain-specific knowledge. Yet, it is superior to manual feature engineering in terms of attack effectiveness. State-of-the-art attacks include Deep Fingerprinting by Sirinam et al.~\cite{df} and Robust Fingerprinting by Shen et al.~\cite{rf}, which both made notable gains compared to k-fingerprinting and CUMUL on mature WF defenses.

\subsubsection{Defenses}
\label{sec:background:wf:defenses}

The community has been working on defenses in parallel to the development of WF attacks. They can be categorized into imitation, regulation, alteration, traffic splitting, and adversarial techniques~\cite{regulator,detorrent}. We briefly describe each category to showcase the diversity of defenses and highlight some notable examples.

\textbf{Imitation} defenses aim to make traffic from visiting one website (nearly) identical to that of another website. Some canonical examples are Walkie-Talkie~\cite{wt}, which groups ``sensitive'' pages with ``non-sensitive'' pages, ensuring each pair has identical burst sequences; and Glove~\cite{glove} and Supersequence~\cite{supersequence}, which group pages into explicit sets and shape their traces to appear identical. A recent defense, Palette~\cite{shen2024real}, uses explicit sets optimized to reduce overhead. Such defenses require knowledge of traces (or their key features) to shape traffic towards, which is a practical concern.

\textbf{Regulation} defenses aim to regulate traffic, making all traces look the same as some target trace or pattern. Tamaraw~\cite{Tamaraw} is a highly effective constant-rate defense but incurs high overheads. WTF-PAD~\cite{wtfpad} is based on the concept of Adaptive Padding~\cite{ap} and attempts to regulate bursts to a target distribution using padding. While WTF-PAD is effective against attacks with manual feature engineering, deep learning-based attacks significantly reduce its effectiveness~\cite{df}. RegulaTor~\cite{regulator} primarily uses padding with minimal latency (blocking) to withstand deep learning-based attacks with modest overheads. Suarakav~\cite{surakav} uses Generative Adversarial Networks (GANs) to generate realistic traffic patterns dynamically and regulates traces to match a generated pattern, similarly with modest overheads.

\textbf{Alteration} defenses aim to alter traffic to such an extent that traces become useless for classification. FRONT~\cite{front} and DeTorrent~\cite{detorrent} are the most effective alteration defenses. FRONT obfuscates the front of a trace with padding, while DeTorrent uses a Long Short-Term Memory (LSTM) network to determine how to add download padding and keeps a fixed 1:5 ratio between upload and download traffic. Interestingly, both are highly randomized padding-only defenses.

\textbf{Traffic splitting} defenses build upon the assumption that the client has multiple paths available and that only one of the paths is observable to an attacker. Which path is observable by the attacker is unknown, and the attacker cannot adaptively choose a path. Defenses unpredictably split traffic between the paths according to some splitting strategy. Three such defenses are TrafficSliver~\cite{trafficsliver}, HyWF~\cite{hywf}, and CoMPS~\cite{comps}. These defenses are less effective against tailored deep learning attacks, even with the niche and perhaps unrealistic assumption that the attacker can only observe a single path~\cite{maturesc}.

\textbf{Adversarial techniques} are used by Mockingbird~\cite{mockingbird} and BLANKET~\cite{blanket} to defend against deep learning-based attacks in particular. While promising, this family of defenses typically struggles against an adaptive attacker~\cite{detorrent}.

\subsection{Real-World Impact of Traffic Analysis}
\label{sec:background:impact}

The real-world impact of traffic analysis attacks has been widely discussed---not only in the website fingerprinting community~\cite{JuarezAADG14,onlinewf,perryCrit,onrealisticwf,cumul} but also in the broader traffic analysis community concerning end-to-end flow correlation/confirmation attacks~\cite{tor, deepcorr, deepcoffea, traceoddity,Danezis04,MurdochD05,KedoganAP02,trilemma,trilemma2}.

In summary, when evaluating defenses, we consider empowered adversaries that may be given unrealistic capabilities or operate under simplified assumptions. Examples include being able to determine the start and end of network traces~\cite{onrealisticwf}, a limited number of possible websites to evaluate~\cite{JuarezAADG14,cumul,wfwo}, and the ability to train under the same network conditions as victims~\cite{onlinewf}. While such empowered attacker models are useful for evaluating defenses, they may not accurately convey the threat posed by traffic analysis~\cite{jansen2024measurement}.

Regardless, a key takeaway is that there is an inherent trade-off between effectiveness and efficiency in the space~\cite{trilemma, trilemma2}, i.e., stronger protection comes at the cost of higher overhead. Being able to \emph{tune} defenses to assessed needs for different use cases is a key feature of any defense (framework). This is where Maybenot gets its name: the goal is to enable integrators to shed doubt on the attacker's ability to draw conclusions from traffic analysis of a protocol's traffic.

%% file: src/framework.tex
Maybenot is a framework for traffic analysis defenses. Defenses are implemented as Maybenot \emph{machines} running in the framework; this approach is adopted from Tor's Circuit Padding Framework~\cite{circpad,circpaddoc}, further described in Section~\ref{sec:related}.  In this section, we look at the Maybenot framework from a top-down perspective, showing how Maybenot can be integrated into a protocol such as TLS or QUIC. Maybenot is written in the Rust programming language, so we use Rust code snippets to complement our explanations.

Maybenot is designed to be used as a library (Rust crate), integrated into existing protocols. There should be one instance of the framework per connection and party/participant. For example, in the case of Tor, this could be per circuit between a client and relay or between relays in the network. Similarly, it could be per connection for TLS, per stream for QUIC, or even per participant in an MLS group chat. To fit such a wide range of protocols, we emphasize that Maybenot is a piece of the puzzle, not a complete solution.

\subsection{Creating an Instance}
\label{sec:framework:new}
Figure~\ref{fig:framework:new} shows how to create an instance of the Maybenot framework. The framework takes zero or more machines (see Section~\ref{sec:machines}) as input that are run in parallel. Creating the framework is a lightweight operation: simply recreate the framework if a machine should be added or removed.

The next two arguments are fractions, i.e., they can have values in the $[0, 1]$ range. They are upper limits on padding traffic and blocking duration, respectively, enforced by the framework for all machines (0 means no limit). \texttt{max\_padding\_frac} is expressed in relation to the total number of packets \textit{queued} on the defended connection so far, and \texttt{max\_blocking\_frac} is relative to the time elapsed since the framework instance was created. Having framework-wide limits is important when multiple machines are used. It also provides a simple way to control the overhead of using Maybenot, potentially as an easy way to tune defenses.

The \texttt{current\_time} is passed to the framework to control the perceived time of the machines. This makes it easier to implement some types of defenses (varying notions of time, e.g. real-world or epoch/step-based) and support the simulation of Maybenot (see Section~\ref{sec:simulator}). It also serves to ease integration by minimizing the number of dependencies in the framework. Note that instants in time and durations are represented as Rust traits (similar to interfaces) in Maybenot. This allows using Maybenot with custom time sources, such as coarsetime\footnote{\url{https://crates.io/crates/coarsetime}}, significantly improving performance on some architectures and for some use cases (e.g. frequent sampling of time from user space requiring system calls).

Finally, \texttt{rng} specifies the PRNG the framework will use for sampling. Exposing the PRNG provides additional flexibility to the integrator and simplifies testing. One particularly noteworthy use case is shared randomness between communicating endpoints using the framework (perhaps based on some protocol-dependent secret), which may be helpful for randomized defenses that require synchronization. It can also contribute to enabling endpoints to reason about the expected behavior of the other endpoint, i.e., whether the observed padding or blocking behavior is expected or not.

\begin{figure}[ht!]
    \centering
    \begin{lstlisting}[language=Rust, style=boxed]
pub fn new(
    machines: AsRef<[Machine]>,
    max_padding_frac: f64,
    max_blocking_frac: f64,
    current_time: Instant,
    rng: RngCore,
) -> Result<Self, Error>\end{lstlisting}
    \caption{Creating an instance of the Maybenot framework.}
    \label{fig:framework:new}
\end{figure}

With an instance of the framework created (or a suitable error returned, e.g. if a fraction is not within the range $[0, 1]$), there are only two more steps to integrate Maybenot into a protocol. Note that the created instance is lightweight and separates machine definitions from their runtime state, so machines can be efficiently shared between many instances.

\subsection{Triggering Events}
\label{sec:framework:events}
Events related to the connection (tunnel) the instance of the framework is protecting need to be reported. The events to trigger are defined in Figure~\ref{fig:framework:events}. In many cases, these events are caused directly or indirectly by the actions of machines in the framework, as explained in Section~\ref{sec:framework:actions}.

When normal (non-padding) or padding data is queued for sending (before encryption, \texttt{NormalSent}/\texttt{PaddingSent}) or has been processed from a packet (after decryption, \texttt{NormalRecv}/\texttt{PaddingRecv}), the framework needs to know. The same applies when an encrypted packet is sent or received via the tunnel (\texttt{TunnelSent}/\texttt{TunnelRecv}). Note that when padding is queued the framework requires the identifier \texttt{MachineId} of the machine that caused the padding.

When a machine starts blocking outgoing traffic, the framework needs the \texttt{MachineId} of the machine (\texttt{BlockingBegin}) and notification when blocking ends (\texttt{BlockingEnd}).

Finally, the integrator must keep track of two timers for each machine, one of which (the \textit{internal} timer) can be set explicitly by the machine. The \texttt{TimerBegin} and \texttt{TimerEnd} events must be triggered when this timer is set and expires, respectively. We explain this in more detail shortly.

\begin{figure}[ht!]
    \centering
    \begin{lstlisting}[language=Rust, style=boxed]
pub enum TriggerEvent {
    NormalRecv,
    PaddingRecv,
    TunnelRecv,
    NormalSent,
    PaddingSent { machine: MachineId },
    TunnelSent,
    BlockingBegin { machine: MachineId },
    BlockingEnd,
    TimerBegin { machine: MachineId },
    TimerEnd { machine: MachineId },
}\end{lstlisting}
    \caption{Events to trigger in the Maybenot framework.}
    \label{fig:framework:events}
\end{figure}

For each event in Figure~\ref{fig:framework:events}, the integrator needs to detect when they occur and trigger them in the framework. This is done by calling the \texttt{trigger\_events} function, as shown in Figure~\ref{fig:framework:trigger}, with one or more events.

\begin{figure}[ht!]
    \centering
    \begin{lstlisting}[language=Rust, style=boxed]
let events = [TriggerEvent::NormalSent];

for action in f.trigger_events(&events, Instant::now())\end{lstlisting}
    \caption{Triggering events in the Maybenot framework.}
    \label{fig:framework:trigger}
\end{figure}

Note that in Figure~\ref{fig:framework:trigger}, the current time is an argument: \texttt{Instant::now()}. As described above, the integrator is responsible for providing a notion of time to the framework.

\subsection{Scheduling and Performing Actions}
\label{sec:framework:actions}

The \texttt{trigger\_events} function in Figure~\ref{fig:framework:trigger} returns an iterator of actions. This is the heart of the integration with the framework: given that one or more events happened at a particular point in time on the connection, which actions should be taken, according to the running machines?

The possible actions are defined in Figure~\ref{fig:framework:actions}. Each machine running in the framework has at most one \textit{scheduled} action (padding or blocking), and other actions are carried out immediately. As an integrator, you need a per-machine timer for scheduling actions and an \textit{internal} timer that the machine can set explicitly (two timers per machine). The integrator is responsible for scheduling and performing the actions returned by the framework as specified. This division of responsibilities is intentional. It allows the framework to remain simple and lightweight while integration can be tailored to the specific protocol and implementation: for example, in Rust, the choice of asynchronous runtime (e.g. Tokio\footnote{\url{https://tokio.rs/}} or async-std\footnote{\url{https://async.rs/}}) is a significant implementation detail.

\begin{figure}[ht!]
    \centering
    \begin{lstlisting}[language=Rust, style=boxed]
pub enum TriggerAction {
    Cancel { machine: MachineId, timer: Timer },
    SendPadding {
        timeout: Duration,
        bypass: bool,
        replace: bool,
        machine: MachineId,
    },
    BlockOutgoing {
        timeout: Duration,
        duration: Duration,
        bypass: bool,
        replace: bool,
        machine: MachineId,
    },
    UpdateTimer {
        duration: Duration,
        replace: bool,
        machine: MachineId,
    },
}\end{lstlisting}
    \caption{Actions returned by the Maybenot framework.}
    \label{fig:framework:actions}
\end{figure}

\subsubsection{Timers}

An integrator must maintain two timers for each \texttt{MachineId}: an \textit{action} timer and an \textit{internal} timer, as shown in Figure~\ref{fig:framework:timers}.

\begin{figure}[ht!]
    \centering
    \begin{lstlisting}[language=Rust, style=boxed]
pub enum Timer {
    Action,
    Internal,
    All,
}\end{lstlisting}
    \caption{Types of timers in the Maybenot framework.}
    \label{fig:framework:timers}
\end{figure}

The action timer is set according to the \texttt{timeout} field of padding and blocking actions (further described shortly). Upon its expiration, the scheduled action for the machine is to be taken. Thus, a \texttt{Cancel} action that specifies the \texttt{Action} timer MUST have the effect of canceling any scheduled action for the machine with the given identifier \texttt{machine}.

Each machine also has an internal timer: though it is not actually internal to the framework, it is updated based on the machine's internal logic. The \texttt{UpdateTimer} action sets the internal timer for the machine with identifier \texttt{machine}. If the \texttt{replace} flag is true, the timer MUST be set with the given \texttt{duration}; otherwise, it MUST be set to the longest of the \textit{remaining} timer duration and the provided duration. The internal timer can be silently canceled via a \texttt{Cancel} action, in which case the \texttt{TimerEnd} event MUST NOT be triggered.

If a \texttt{Cancel} action has timer \texttt{All}, both the scheduled action for the machine and its internal timer MUST be canceled.

\subsubsection{Send Padding}

Schedule sending a padding packet after a given \texttt{timeout}. The padding size is not specified; this is left to the integrator to decide. Two flags control how the padding is sent: \texttt{bypass} and \texttt{replace}.

The \texttt{bypass} flag indicates if the padding packet MUST be sent despite active blocking of outgoing traffic (``bypass blocking''). This is only allowed if the active blocking was set with the \texttt{bypass} flag (``bypassable blocking''); see Section~\ref{sec:framework:block}. If the \texttt{replace} flag is set, it indicates that the padding packet MUST be replaced by any enqueued (and blocked from being sent on the tunnel) normal data, if available.

The combination of \texttt{bypass} and \texttt{replace} flags can be used to build machines that send constant-rate traffic. For example, consider a machine that first triggers bypassable blocking and then, at a constant rate, triggers the action to send padding with the \texttt{bypass} and \texttt{replace} flags set. At each action, either padding or enqueued normal data will be sent.

\subsubsection{Block Outgoing}
\label{sec:framework:block}

Schedule the blocking of outgoing traffic for a \texttt{duration} after a given \texttt{timeout}. The blocking is for all outgoing traffic on the connection across all machines running in the framework instance. Two flags control how the blocking is performed: \texttt{bypass} and \texttt{replace}.

If the \texttt{bypass} flag is set, the caused blocking MAY be bypassed by padding packets with the \texttt{bypass} flag set. We call this ``bypassable blocking''. If the \texttt{bypass} flag is not set, the caused blocking MUST NOT be bypassed by any means. This design is motivated by ensuring that it is possible to create fail-closed blocking defenses.

If the \texttt{replace} flag is set, the blocking duration MUST replace any current active blocking duration. If the \texttt{replace} flag is not set, the new blocking duration MUST be the longest of the \textit{remaining} blocking duration and the provided duration. This ensures that the only way to reduce the blocking duration is by setting the \texttt{replace} flag. Any bypass flag on active blocking is only updated if the duration is updated (i.e., longer duration or the \texttt{replace} flag is set).

\subsection{Example Main Loop}
Figure~\ref{fig:framework:example} shows an example main loop to illustrate the framework's intended usage. In gist, select your machines and create an instance of the framework. Then, in a loop, periodically collect events and trigger them in the framework. The framework will return an iterator of actions to schedule. The integrator is responsible for performing the actions as specified.

\begin{figure}[ht!]
    \centering
    \begin{lstlisting}[language=Rust, style=boxed]
use maybenot::{
    Framework,
    Machine,
    TriggerAction,
    TriggerEvent,
};
use std::{str::FromStr, time::Instant};

// parse a machine from a string
let s = "02eNpjYEAHjOgCAAA0AAI=";
let m = vec![Machine::from_str(s).unwrap()];

// create an instance of the framework
let mut f = let mut f = Framework::new(&m, 0.0, 0.0, Instant::now(), rand::thread_rng()).unwrap();

loop {
    // collect one or more events
    let events = [TriggerEvent::NormalSent];

    // trigger events and get actions
    for action in f.trigger_events(&events, Instant::now()) {
        match action {
            TriggerAction::Cancel {
                machine: _,
                timer: _,
            } => {
                // cancel timer
            }
            TriggerAction::SendPadding {
                timeout: _,
                bypass: _,
                replace: _,
                machine: _,
            } => {
                // send padding
            }
            TriggerAction::BlockOutgoing {
                timeout: _,
                duration: _,
                bypass: _,
                replace: _,
                machine: _,
            } => {
                // block outgoing traffic
            }
            TriggerAction::UpdateTimer {
                duration: _,
                replace: _,
                machine: _,
            } => {
                // set internal timer
            }
        }
    }
}\end{lstlisting}
    \caption{Example main loop.}
    \label{fig:framework:example}
    \vspace{-6pt}
\end{figure}

Note that \texttt{trigger\_events} takes a slice of events as input, making it possible to trigger multiple events simultaneously. Regardless of the number of events, the result is a maximum of one action per machine running in the framework. This skips unnecessary updates to actions in case the main loop is run infrequently or the number of events is high. Though ideally there would be a separate loop iteration for each event, as a rule of thumb, drop/overwrite the oldest events first if you run out of space for events to trigger (e.g. a full events queue).

%% file: src/machines.tex
The Maybenot framework is a runtime for Maybenot machines. A machine has some runtime state in the framework, processes triggered events, and produces actions. Each machine has up to one scheduled action at any point in time.

Maybenot machines are probabilistic finite-state machines. They are based on the notion of ``padding machines'' (non-deterministic finite-state machines) in the Tor Circuit Padding Framework~\cite{circpad,circpaddoc}, further discussed in Section~\ref{sec:related}.

To explain Maybenot machines, we take a bottom-up approach, starting with distributions (Section~\ref{sec:machines:dist}), then states (Section~\ref{sec:machines:state}), and finally machines (Section~\ref{sec:machines:machine}).

\subsection{Distributions}
\label{sec:machines:dist}
Central to a machine is the notion of a distribution. Distributions are frequently sampled as machines transition between states and produce actions. Figure~\ref{fig:machines:distlist} shows the types of distributions supported by Maybenot machines, provided by the rand\_distr\footnote{\url{https://crates.io/crates/rand\_distr}} crate. Likely, the supported distributions are both too many and wrong for many applications. Extending support is straightforward, though.

\begin{figure}[ht!]
    \centering
    \begin{lstlisting}[language=Rust, style=boxed]
pub enum DistType {
    Uniform { low: f64, high: f64 },
    Normal { mean: f64, stdev: f64 },
    SkewNormal { location: f64, scale: f64, shape: f64 },
    LogNormal { mu: f64, sigma: f64 },
    Binomial { trials: u64, probability: f64 },
    Geometric { probability: f64 },
    Pareto { scale: f64, shape: f64 },
    Poisson { lambda: f64 },
    Weibull { scale: f64, shape: f64 },
    Gamma { scale: f64, shape: f64 },
    Beta { alpha: f64, beta: f64 },
}
\end{lstlisting}
    \caption{Supported distributions.}
    \label{fig:machines:distlist}
\end{figure}

Figure~\ref{fig:machines:diststruct} shows the core \texttt{Dist} struct and the \texttt{sample} method used to sample distributions. The \texttt{dist} field specifies the distribution type and parameters. The \texttt{start} field is the starting value added to the value sampled from the distribution, and \texttt{max} is the maximum value that can be sampled (0 for no limit). Note that, due to what the sampled values are used for (explained in Section~\ref{sec:machines:state}), the resulting value is clamped to be $[0.0, $max$]$. Values are floats, rounded to discrete values if needed based on what they will be used for. When sampling times, the framework operates in microseconds.

\begin{figure}[ht!]
    \centering
    \begin{lstlisting}[language=Rust, style=boxed]
pub struct Dist {
    pub dist: DistType,
    pub start: f64,
    pub max: f64,
}

// impl Dist
pub fn sample(self, rng: &mut R) -> f64 {
        let mut r: f64 = 0.0;
        r = r.max(self.dist_sample(rng) + self.start);
        if self.max > 0.0 {
            return r.min(self.max);
        }
        r
}
\end{lstlisting}
    \caption{Sampling distributions.}
    \label{fig:machines:diststruct}
\end{figure}

Sampling distributions is the most computationally expensive part of triggering events in the Maybenot framework. Efficient PRNGs may significantly improve performance, but, to err on the side of caution, we recommend using a cryptographically secure PRNG. Interesting future work is thoroughly evaluating if using a non-cryptographically secure PRNG would be safe for some use cases (famous last words).

\begin{figure}[ht!]
    \centering
    \begin{lstlisting}[language=Rust, style=boxed]
pub struct State {
    pub action: Option<Action>,
    pub counter: (Option<Counter>, Option<Counter>),
    // private field: transitions
}
\end{lstlisting}
    \caption{States that make up a machine.}
    \label{fig:machines:state}
\end{figure}

\subsection{States}
\label{sec:machines:state}

Figure~\ref{fig:machines:state} shows the core \texttt{State} struct of machines. It consists of an action, a counter update, and state transitions.

\subsubsection{Actions}
\label{sec:machines:state:action}
A state can either be a no-op state by setting the \texttt{action} field to \texttt{None}, or it can optionally specify one of the four actions in Figure~\ref{fig:machines:action}.

\begin{figure}[ht!]
    \centering
    \begin{lstlisting}[language=Rust, style=boxed]
pub enum Action {
    Cancel { timer: Timer },
    SendPadding {
        bypass: bool,
        replace: bool,
        timeout: Dist,
        limit: Option<Dist>,
    },
    BlockOutgoing {
        bypass: bool,
        replace: bool,
        timeout: Dist,
        duration: Dist,
        limit: Option<Dist>,
    },
    UpdateTimer {
        replace: bool,
        duration: Dist,
        limit: Option<Dist>,
    },
}
\end{lstlisting}
    \caption{Actions associated with a state.}
    \label{fig:machines:action}
\end{figure}

An action has associated distributions that are sampled. The \texttt{duration} distributions for blocking and timer actions are used to sample the duration that will be returned to the integrator. The \texttt{bypass} and \texttt{replace} flags are used to carry out the actions as described in Section~\ref{sec:framework:actions}. An action is scheduled every time a machine transitions. Note that, since only one action is allowed per machine per call to \texttt{trigger\_events}, the action of the last state (with an action: note from Figure~\ref{fig:machines:state} that the action is optional) transitioned to is scheduled.

\subsubsection{Counters}
\label{sec:machines:state:counter}

Each machine has two (discrete) counters, updated upon transition to a state if its \texttt{counter} field is set. The \texttt{Counter} struct, shown in Figure~\ref{fig:machines:counter}, specifies the operation to apply to a counter (increment, decrement, or set), an optional distribution to sample a value from (default one if not set), and if a copy should occur, in which case the previous value of the other counter is used and the distribution is ignored. When either counter is decremented to zero, the framework triggers a \texttt{CounterZero} event internally for the machine. However, to prevent infinite looping and restrict execution time, only one \texttt{CounterZero} event may be triggered per counter and call to \texttt{trigger\_events} across all machines. The counters are part of the machine runtime kept by the framework.

\begin{figure}[ht!]
    \centering
    \begin{lstlisting}[language=Rust, style=boxed]
pub enum Operation { Increment, Decrement, Set }

pub struct Counter {
    pub operation: Operation,
    pub dist: Option<Dist>,
    pub copy: bool,
}
\end{lstlisting}
    \caption{Counter updates associated with a state.}
    \label{fig:machines:counter}
    \vspace{-4pt}
\end{figure}

\subsubsection{Transitions}
\label{sec:machines:state:transitions}
Each state has a vector of state transitions for all possible events (Figure~\ref{fig:machines:event}). State transitions are specified with \texttt{Trans}, a tuple struct consisting of a state index to transition to and the probability. Probabilities must sum to at most one. Figure~\ref{fig:machines:transition} shows an example with transitions on four events. Observe that the probabilities do not need to sum to one. This is to support machines that only transition with a small probability.

The state index may also be the pseudo-state \texttt{STATE\_END}, which does not cancel any scheduled timer but permanently ends the machine, preventing future transitions. Another pseudo-state, \texttt{STATE\_SIGNAL}, sends a signal to all other machines running in the framework---i.e., transitioning to it triggers a \texttt{Signal} event. If multiple machines signal during the same call to \texttt{trigger\_events} (including when responding to signals), all machines receive exactly one signal.

\begin{figure}[ht!]
    \centering
    \begin{lstlisting}[language=Rust, style=boxed]
pub enum Event {
    NormalRecv,
    PaddingRecv,
    TunnelRecv,
    NormalSent,
    PaddingSent,
    TunnelSent,
    BlockingBegin,
    BlockingEnd,
    LimitReached,
    CounterZero,
    TimerBegin,
    TimerEnd,
    Signal,
}
\end{lstlisting}
    \caption{Events for state transitions.}
    \label{fig:machines:event}
\end{figure}

\begin{figure}[ht!]
    \centering
    \begin{lstlisting}[language=Rust, style=boxed]
let first_state = State::new(enum_map! {
    Event::PaddingSent => vec![Trans(1, 1.0)],
    Event::NormalSent => vec![Trans(2, 0.7)],
    Event::NormalRecv => vec![Trans(1, 0.6), Trans(2, 0.4)],
    Event::PaddingRecv => vec![Trans(STATE_END, 1.0)],
_ => vec![],
});
\end{lstlisting}
    \caption{Example of creating a new \texttt{State} with transitions.}
    \label{fig:machines:transition}
    \vspace{-8pt}
\end{figure}

\subsubsection{Limits}
\label{sec:machines:state:limit}
A state can transition to itself, and a (discrete) limit can be sampled on the number of actions allowed to be repeatedly scheduled due to such self-transitions. There is no limit if the \texttt{limit} dist of the state's action is \texttt{None}. Otherwise, the limit is decremented when padding is queued, blocking is started, or the internal timer is set. When the limit reaches zero, the framework triggers a \texttt{LimitReached} event internally for the corresponding machine. A new limit is sampled each time a state is transitioned into from another state.

\subsection{Machines}
\label{sec:machines:machine}
Figure \ref{fig:machines:machine} shows the \texttt{Machine} struct. In gist, a machine consists of a vector of states and fields that limit the machine's behavior. When a machine is added to the Maybenot framework, it is set to state 0 and no action is scheduled. Actions are scheduled when/if the machine transitions. Note that a limit is sampled for state 0 to allow immediate self-transitions.

\begin{figure}[ht!]
    \centering
    \begin{lstlisting}[language=Rust, style=boxed]
pub struct Machine {
    pub allowed_padding_packets: u64,
    pub max_padding_frac: f64,
    pub allowed_blocked_microsec: u64,
    pub max_blocking_frac: f64,
    pub states: Vec<State>,
}
\end{lstlisting}
    \caption{A Maybenot machine.}
    \label{fig:machines:machine}
    \vspace{-8pt}
\end{figure}

\subsubsection{Limits and Allowed Actions}
\label{sec:machines:limits}
Per-machine limits are set by the \texttt{max\_padding\_frac} and \texttt{max\_blocking\_frac} fields. This is in addition to state limits (Section~\ref{sec:machines:state:limit}) and global framework limits (Section~\ref{sec:framework:new}).  If any limit is hit, the machine will be prevented from scheduling an action on state transitions until below all the limits.

However, the two fields \texttt{allowed\_padding\_packets} and \texttt{allowed\_blocked\_microsec} are not limits but rather budgets of allowed actions. These budgets ignore all limits and should, therefore, be used carefully. This design might seem a bit reckless, but it is motivated by machines that act early on in the connection, when any reasonable fraction of limits would be exceeded. For example, a machine might want to obfuscate handshakes, as is done in Tor for the setup of onion circuits~\cite{circpad}. Machines are given their budgets on framework instance creation and all actions (regardless of if allowed by limits or not) subtract from them until depleted.

\subsubsection{Serialization and Deserialization}
Machines can be serialized to and deserialized from Base64-encoded strings. We use Serde\footnote{\url{https://serde.rs/}}, a commonly used framework for serialization and deserialization in Rust, together with the bincode\footnote{\url{https://github.com/bincode-org/bincode}} encoder/decoder. Encoded machines are then compressed using zlib\footnote{\url{https://www.zlib.net/}} before finally being Base64-encoded. By default, Maybenot enforces a maximum \emph{decompressed} memory size of 1 MiB and performs extensive parameter validation to make it safer to parse machines from untrusted inputs. The use of Serde makes it trivial to create custom serialization and deserialization logic if needed.

Note that a machine is separate from its runtime (memory keeping track of its current state) kept in each instance of the framework using it. This means that many instances of the framework can share the same machine, since each instance will have its own runtime. No runtime information is ever serialized. The size of the runtime is constant, i.e., the number of states in a machine has no impact on runtime memory.

%% file: src/simulator.tex
Developing effective and efficient machines is challenging. In an ideal world, machines would be evaluated by collecting real-world datasets of the machines running in Maybenot integrated with the intended protocol. Unfortunately, this is typically prohibitively expensive. For such cases, we provide a simple bare-bones simulator implemented in Rust as a crate, similar to the framework.

The goal of the simulator is not to be perfect---whatever that would entail, given that the framework is designed to be integrated into a wide range of protocols---but to be useful. Hopefully, most development can be done with the simulator and only fine-tuning is needed for later integration.

The core idea of the simulator is to simulate how a base network trace would have looked if one or more machines were running at the client and/or server. Therefore, there are two steps to using the simulator: parsing a base network trace and simulating machines on the trace.

\subsection{Parsing a Base Network Trace}
Figure~\ref{fig:simulator:trace} shows a network trace with ten packets when visiting a website and how to parse it. The trace is a string of lines, where each line is a packet with the format ``timestamp,direction\textbackslash n''. The timestamp is the number of nanoseconds since the start of the trace, and the direction is either ``s'' for sent or ``r'' for received (from the client's perspective).

To parse the trace, the simulator also takes a network model of the latency between the client and server. The network model is used to simulate \emph{event queues} for the client and server, so packets are sent and arrive at the client precisely as in the provided trace. This is a crude approximation of the network between the client and server and should be improved to make the simulator more useful in the long term.

\begin{figure}[ht!]
    \centering
    \begin{lstlisting}[language=Rust, style=boxed]
use maybenot_simulator::{network::Network, parse_trace};
use std::time::Duration;

let raw_trace = "0,s
19714282,r
183976147,s
243699564,r
1696037773,s
2047985926,s
2055955094,r
9401039609,s
9401094589,s
9420892765,r";

let network = Network::new(Duration::from_millis(10), None);
let mut input_trace = parse_trace(raw_trace, &network);\end{lstlisting}
    \caption{Parsing an example trace.}
    \label{fig:simulator:trace}
\end{figure}

The resulting input trace is a queue of Maybenot events. Note that the simulator will change the queue (``mut''), so repeated simulations using the same queue must clone it.

\subsection{Simulating Machines}
Figure~\ref{fig:simulator:sim} shows an example of simulating a machine on the trace from Figure~\ref{fig:simulator:trace}. The simulator supports zero or more machines running at the client and server. Because machines may run forever (e.g. sending more padding on padding being sent), it is possible to set the maximum number of events (client and server) to simulate and a flag to filter out events only related to network packets. The simulator's output is a vector of events describing the simulated trace.

\begin{figure}[ht!]
    \centering
    \begin{lstlisting}[language=Rust, style=boxed]
use maybenot_simulator::sim;
use maybenot::machine::Machine;
use std::{str::FromStr, time::Duration};

let s = "02eNp1ibEJAEAIA5...";
let m = Machine::from_str(s).unwrap();
        
let trace = sim(
    &[m], // client machines
    &[], // server machines
    &mut input_trace,
    network.delay,
    100, // max events
    true, // only packet events?
);
\end{lstlisting}
    \caption{Simulating a machine on a trace.}
    \label{fig:simulator:sim}
\end{figure}

Parsing the output is straightforward. The output is a vector of Maybenot events, including a simulated timestamp, a flag indicating whether the event is from the client or server, and a flag (for \texttt{TunnelSent} and \texttt{TunnelRecv} events) indicating whether the packet contains padding. Figure~\ref{fig:simulator:parse} shows an example of printing events related to network packets going through the tunnel at the client.

\begin{figure}[ht!]
    \centering
    \begin{lstlisting}[language=Rust, style=boxed]
let starting_time = trace[0].time;
trace
    .into_iter()
    .filter(|p| p.client)
    .for_each(|p| match p.event {
        TriggerEvent::TunnelSent => {
            if p.contains_padding {
                println!(
                    "sent a padding packet at {} ms",
                    (p.time - starting_time).as_millis()
                );
            } else {
                println!(
                    "sent a normal packet at {} ms",
                    (p.time - starting_time).as_millis()
                );
            }
        }
        TriggerEvent::TunnelRecv => {
            if p.contains_padding {
                println!(
                    "received a padding packet at {} ms",
                    (p.time - starting_time).as_millis()
                );
            } else {
                println!(
                    "received a normal packet at {} ms",
                    (p.time - starting_time).as_millis()
                );
            }
        }
        _ => {}
    });\end{lstlisting}
    \caption{Printing packet-related events at the client.}
    \label{fig:simulator:parse}
\end{figure}

For more advanced use cases, the simulator exposes a \texttt{sim\_advanced()} method with more comprehensive settings; see Figure~\ref{fig:simulator:args}. Notably, the simulator supports optional integration delays at the client and server. Integration delays, as the name suggests, can be present in integrations of Maybenot, for example, if Maybenot is running in user space and an integrated protocol runs in kernel space. The simulator supports delays surrounding performing actions and reporting events; see the Rust documentation for more details.

\begin{figure}[ht!]
    \centering
    \begin{lstlisting}[language=Rust, style=boxed]
/// Arguments for [`sim_advanced`].
#[derive(Clone, Debug)]
pub struct SimulatorArgs<'a> {
    pub network: &'a Network,
    pub max_trace_length: usize,
    pub max_sim_iterations: usize,
    pub only_client_events: bool,
    pub only_network_activity: bool,
    pub max_padding_frac_client: f64,
    pub max_blocking_frac_client: f64,
    pub max_padding_frac_server: f64,
    pub max_blocking_frac_server: f64,
    pub insecure_rng_seed: Option<u64>,
    pub client_integration: Option<&'a Integration>,
    pub server_integration: Option<&'a Integration>,
}\end{lstlisting}
    \caption{Advanced simulator arguments.}
    \label{fig:simulator:args}
\end{figure}

\hfill
\\

\subsection{Simulator Internals}
One detail of the internals of the simulator to highlight is its core logic in moving simulated time forward. The internal state of the simulator is driven by four possible sources for the next event to simulate:

\begin{description}
    \item[event queue] the next event in the main event queue given the state of
    the client and server (active blocking, bypass and replace flags, etc).
    \item[block expiry] active blocking expiring at the client or server.
    \item[internal timers] an internal timer for a machine expires.
    \item[scheduled actions] any action scheduled by a machine running at the
    client or server.
\end{description}
The simulator will always simulate the next event from one of these sources. When sources have their events simultaneously, the simulator will prioritize in the order of event queue, block expiry, internal timers, and, finally, scheduled actions. The logic behind this is that events in the event queue typically happen outside of Maybenot (e.g. network packets), while blocking expiration, internal timers, and scheduled actions are tied to how Maybenot is integrated into a protocol. We assume that such logic will not be integrated inline in the protocol but instead run in a lightweight thread or similar. The above prioritization in the simulator might be noticed when base network traces lack timestamp granularity and many resulting events end up at the same time in the parsed queue.

%% file: src/discussion.tex
The ultimate problem at hand---improved traffic analysis driven by rapid developments in AI/ML---is shared by many protocols. So are the key building blocks of defenses: padding and blocking. Maybenot is a library that aims to capture the main common features of traffic analysis defenses and leave the protocol-specific integration to the integrator, e.g. negotiation and how to cause the padding and blocking. The hope is that this will allow for tailored use of the library in different protocols, with the added benefit of a common framework with many available defenses. What features should or should not be included in the framework is a balancing act.

\subsection{Padding and Chaff}
One important feature missing from Maybenot is inherited from the focus on Tor with constant-size packets (cells) in the defense community. Maybenot does not address the question of padding packets to fixed sizes but instead adopts the simple notion of padding packets (chaff). We believe it is an open question whether it is even possible to make an effective and efficient defense against more advanced traffic analysis, such as website fingerprinting, on normal packets of variable sizes. As part of integrating Maybenot into a new protocol, the integrator will likely need to carefully consider this issue in their analysis.

\subsection{Sending and Queues}
Version 2 of Maybenot provides events for queuing/processing normal or padding data and sending/receiving encrypted packets over the defended tunnel; however, what this means in practice will depend on the details of the protocol with which the framework is integrated.

Consider the \texttt{PaddingSent} event (padding data queued to be sent). That the event is to be triggered as part of performing the scheduled padding action is natural. However, how this is implemented in the protocol is not. For example, it could be implemented as custom (distinguishable) data queued to be sent through the protocol, just like other application data. However, it could also be implemented as a particular packet type in the protocol sent immediately. The differences are significant because in the former case, the padding would be \emph{queued} up with other application data and, in the latter, immediately turned into a packet and \emph{egress queued} (delivered) in the network stack heading towards the NIC.

From the above, it should be clear that the exact semantics of the \texttt{PaddingSent} event are not defined by the framework; the same is the case for the \texttt{NormalSent}, \texttt{TunnelSent}, and receive events. This may or may not matter.  Note that the distinction between queuing (\texttt{PaddingSent} and \texttt{NormalSent}) and delivery (\texttt{TunnelSent}) should, if possible, be made in such a way as to provide benefits during periods of congestion, allowing padding limits to be reached even if packets are dropped or delayed due to network conditions. As a rule of thumb, what ``queueing'' entails will depend on the specific protocol that the framework is integrated with, but it should capture any attempt to send data via the protocol, and we expect ``tunnel'' events to occur when data has been passed to some lower layer in the network stack. In general, machines that make sense with the selected semantics should be chosen.

\subsection{Expressing Defenses}
Another consideration is the focus on state machines, inherited from the Tor Circuit Padding Framework~\cite{circpad,circpaddoc}. While state machines have been used in the community in the past~\cite{ap,circpaddoc,circuitfp} and Maybenot expands on this by supporting multiple probabilistic machines running in parallel, it is far from a given that this is the best approach. Smith et al. in QCSD~\cite{qcsd} use a single regularization algorithm with target traces as input, and Gong et al. in WFDefProxy~\cite{wfdefproxy} simply use the programming language Go. Another example is FAN by Rochet and Elahi~\cite{fan}, which uses eBPF~\cite{ebpf,xdp} to make anonymous networks more flexible by allowing dynamic updates to protocols, including the Tor Circuit Padding Framework. Using eBPF or similar technologies, such as WebAssembly~\cite{wasm}, may allow for richer defenses than the state machines offered by Maybenot while retaining the flexibility gained from not using a hardcoded defense. This comes at the cost of much added complexity, though, compared to the simple state machines in Maybenot with a ``runtime'' of 1,555 loc in Rust (and 2,467 loc of tests) at the time of writing.

Consider Maybenot's API for integration, covered in Section~\ref{sec:framework}. Ignoring the machines for a moment, the main loop is basically  ``report relevant events at a point in time and get actions to schedule in return''. This interface fits the vast majority of defenses. Even the recent work of using GANs (e.g. Surakav~\cite{surakav}) and LSTMs (DeTorrent~\cite{detorrent}) as defenses can be expressed in terms of such an interface. Surakav~\cite{surakav} samples burst sizes from a GAN (discrete steps mapped to bursts), while DeTorrent operates on log-scale timed bins. Future versions of Maybenot could replace state machines with, say, neural networks or more general-purpose programming languages with few API changes as a result. Alternatively, solutions such as Surakav or DeTorrent could be used to generate state machines for Maybenot. One reason to prefer state machines over, e.g., neural networks is performance. Inline inferences in neural networks are typically much slower than state machines, though the very rapid ongoing developments in AI/ML may change this in the long term.

Finally, we remark that version 2 of the framework includes two counters per machine, a per-machine timer, and a mechanism for explicit signaling between machines. We expect this to improve the framework's expressiveness substantially while maintaining its simplicity. However, Maybenot is the first state machine-based framework to introduce such notions; thus, their practical utility for implementing defenses---and whether they sufficiently support effective defenses---is an open question to be answered through experience. This will aid future iterative development of the framework.

\subsection{Why a Framework}
Why bother with a framework for defenses instead of directly implementing defenses? For one, defenses are often moving targets. The last decade has seen large improvements in traffic analysis attacks, particularly around website fingerprinting, and defenses have been improving in response. A prime example here is from the Tor Project. They started out intending to implement the WTF-PAD defense by Ju{\'{a}}rez et al.~\cite{wtfpad}, but the Deep Fingerprinting attack by Sirinam et al.~\cite{df} greatly reduced its effectiveness compared to earlier evaluations against (among others) the k-fingerprinting attack by Hayes and Danezis~\cite{kfp}. So, instead of implementing WTF-PAD, the Tor Circuit Padding Framework was born~\cite{circpad,circpaddoc}.

A framework also allows for the easy combination of multiple defenses. Combinations have been shown to be effective~\cite{hywf,comps,surakav}. The selection of defenses to combine could also be dynamic and adaptive, e.g. based on current normal traffic to hide moments of inactivity or turned off for bulk downloads. A framework is part of \emph{orchestrating} defenses.

Another consideration is moving from \textit{website} fingerprinting to \textit{webpage} fingerprinting defenses. While most web traffic is encrypted, it goes directly between the client and the server. Any network attacker can, therefore, in the vast majority of cases, simply perform website fingerprinting by observing all relevant IP addresses~\cite{wfmasses}. In a webpage fingerprinting setting, optimal defenses would be per website and optimized for the pages distributed by that website. It is known that application-layer knowledge can be used to create more effective and efficient defenses~\cite{wfapplayer}. A framework for defenses integrated into, e.g., QUIC or HTTP/3 would allow for tailored per-site defenses. Websites could distribute serialized defenses to clients upon connection establishment, or the server could implement the client side of the defense partly in the application layer (the inverse of QCSD~\cite{qcsd}).

\subsection{Simulator Limitations}
The simulator has several limitations. For one, it is a simulator! We are simulating the integration with the application/destination using the framework and the network between the client and server/relay. We have a \emph{sim2real} problem.

In terms of networking, we use a fixed static delay. This should be improved and evaluated against real-world network experiments. The goal of the simulator is not necessarily to be a perfect simulator but to be a useful simulator for making different kinds of traffic analysis defenses. One possible source of inspiration for improving the networking is Shadow~\cite{shadow}.

There are also fundamental issues with simulating the blocking actions of machines. Because the simulator takes as input a base network trace of encrypted network traffic, we do not know any semantics or inter-dependencies between the packets in the encrypted trace. As a result, we cannot correctly simulate blocking actions. For example, if a machine blocks a packet, we cannot know if the blocked packet contains a request for a resource that leads to a response in the following received packets. The simulator will still happily receive the resource in the encrypted network trace.

\subsection{Lessons Learned from Version 1}
Early work on implementing defenses in version 1 of Maybenot (e.g.~\cite{maybenotwpes}) yielded several insights into which capabilities are needed to support efficient and effective defenses. In particular, it became clear that it is possible to approximate several hand-crafted defenses using a simple probabilistic state machine model. Still, more expressiveness is required to fully capture their intended behavior and achieve the desired trade-offs between overhead and protection.

It is often necessary to count events--- primarily related to packets sent/received---and this was previously accomplished via machines with many states and carefully chosen limits on self-transitions. However, this can complicate machine design and lead to significant runtime overheads, rendering defenses impractical. Thus, a significant addition to Maybenot v2 is counters: two are available to each machine, and they can optionally be updated upon transition to a state. A \texttt{CounterZero} event is triggered when either is decremented to zero. This raises the grammar of Maybenot in the Chomsky hierarchy (in fact, two counters can, in principle, simulate an arbitrary Turing machine~\cite{wiki-twocounters}) and facilitates simple machine design, richer machines, and efficiency.

Time-based actions are also common in traffic analysis defenses, but it was not possible for machines to explicitly measure time in Maybenot v1. This has led to introducing a per-machine \textit{internal} timer, updated based on a machine's internal logic but maintained by the integrator. The \texttt{UpdateTimer} action can be used to set the internal timer, and \texttt{TimerBegin} and \texttt{TimerEnd} events signal that it has started or expired, respectively. One advantage of having the integrator be responsible for the timer is that varying notions of time (e.g. discrete steps or wall time) can be selected depending on the use case. We expect this feature to be handy when implementing hand-crafted defenses.

In the same fashion, it is now possible for the integrator to specify a PRNG for all sampling done by the framework. Though we caution that using a non-cryptographically secure PRNG may have unintended and unexpected consequences, this allows for selecting the trade-off between efficiency and statistical properties. It also paves the way for randomized defenses that require synchronization: the same seed can be used for the PRNG on both ends of a defended tunnel. An example of a defense that may benefit from this capability is Surakav, which could not be implemented in v1~\cite{maybenotdefenses}.

Even at a more local scale, maintaining a shared state between or otherwise synchronizing multiple machines running in the same framework instance is often desirable. For this, we have added a \texttt{STATE\_SIGNAL} pseudo-state, which, when transitioned to, sends a signal to all other running machines. This supports Maybenot's vision of providing a platform for orchestrating defenses and eliminates the need for, e.g., synchronizing by re-enabling blocking, which is cumbersome, affected by integration delays, and not semantically correct.

In the interest of simplicity, Maybenot v2 no longer supports features that were found to be of little use. As it is unclear whether it is even possible to create effective, efficient defenses in settings with variable packet sizes, support for packet sizes and knowledge of the MTU of the underlying tunnel have been removed in favor of packet counts. As a result, the \texttt{include\_small\_packets} flags for machines have been removed. Similarly, states no longer contain a \texttt{limit\_includes\_nonpadding} flag: limits never include normal packets, as this is a niche feature. Defenders interested in such limits may now opt to use counters instead.

Recognizing the diversity of protocols that may integrate the framework and the commonalities among their mechanics, we have reimagined how packet-related events are reported to the framework. We begin with a slight change in terminology: ``non-padding'' packets (carrying application data) are now referred to as ``normal'' throughout the framework, including in event names. Importantly, it may be useful for certain defenses to differentiate between sending and receiving normal/padding data. In contrast, others may need to know precisely when a packet has been delivered to or received from a lower layer in the network stack. Thus, the \texttt{\{Normal,Padding\}+\{Sent,Recv\}} events now refer to queueing data to be sent or processing it from a packet, while the new \texttt{TunnelSent} and \texttt{TunnelRecv} events will be triggered when data has actually been delivered or received.

We have also endeavored to improve Maybenot's interface. For instance, an explicit \texttt{Cancel} has been added, which allows a machine to cancel a pending action timer, the internal timer, or both; in tandem, the redundant pseudo-state \texttt{STATE\_CANCEL} has been removed, simplifying state transitions. Much of the codebase has also been refactored to improve the interface for creating and modifying machines. Events, actions, and counter updates are now represented by separate data structures linked to states, among other changes.

Finally, some miscellaneous updates include improvements to serialization (Section~\ref{sec:machines:machine}), the addition of the SkewNormal distribution, and an optional \texttt{parsing} feature to retain partial support for v1 machines. The described changes make Maybenot more expressive, efficient, and user-friendly. Yet, much remains to learn: we plan to inform future development based on experiences with v2 integrations.

%% file: src/related.tex
Maybenot is based on the Tor Circuit Padding Framework~\cite{circpad,circpaddoc} by Perry and Kadianakis. The Tor Circuit Padding Framework is, in turn, a generalization of WTF-PAD~\cite{wtfpad} by Ju{\'{a}}rez et al., a website fingerprinting defense based on the concept of Adaptive Padding by Shmatikov and Wang~\cite{ap}.

\subsection{Tor Circuit Padding Framework}
As the name suggests, the Tor Circuit Padding Framework is a framework for implementing padding in Tor circuits. It is closely integrated into Tor's C codebase. Clients can negotiate fixed hardcoded ``padding machines'' with relays in the network. At the time of writing, while the framework is deployed as part of Tor, only two small padding machines are active in the network~\cite{circpad}. Their goal is to hide the setup of client-side onion service circuits, making them produce the same sequence of cells as non-onion circuits.

Comparing Maybenot to the Tor Circuit Padding Framework, the main differences are that Maybenot is designed to be integrated into a wide range of protocols, is written in Rust as a library, and includes a richer set of features. The Tor Circuit Padding Framework is closely tied to Tor and supports negotiation of padding machines, which Maybenot considers out of scope. Maybenot machines support probabilistic transitions, padding, blocking, and associated bypass/replace flags, counters, timers, and signaling. Maybenot also has more distributions but removes support for histograms. Finally, Maybenot does not support RTT-based estimates as offsets for timers. RTT-based estimates and histograms are excluded due to a lack of identified use cases, simplifying Maybenot.

\subsection{QCSD}
QCSD~\cite{qcsd} by Smith et al. is a framework for traffic analysis defenses tailored to QUIC~\cite{quic}. The main focus of QCSD is to be a \emph{client-side} framework, generating padding and delaying traffic at the server by leveraging protocol features of QUIC and HTTP/3~\cite{http3}. This is significant for fostering adoption of defenses, as it removes the need to modify the server or any party other than the client. However, this strength is also QCSD's main weakness: its dependency on QUIC and HTTP/3 at endpoints lacks support for defending only between the client and intermediate relays/proxies. Also, websites are typically made up of resources from multiple domains/endpoints, adding further complexities.

The analog to Maybenot machines in QCSD is a regularization algorithm running at the client, which shapes the connection according to a provided target trace (static or dynamically generated). In principle, the regularization algorithm should be possible to implement as Maybenot machines are generated based on the target trace. 
Shaping traffic from the server to the client involves control messages, which adds some overhead and makes it challenging to ensure exact shaping at the server. Smith et al. highlight that extensions to QUIC may assist clients in precisely shaping server traffic.

\subsection{WFDefProxy}
WFDefProxy~\cite{wfdefproxy} by Gong et al. is a platform for implementing website fingerprinting defenses and empirically evaluating them in real networks. WFDefProxy is based on obfs4~\cite{obfs4}, a Pluggable Transport (PT)~\cite{pt} for Tor. Being based on obfs4 as a PT, WFDefProxy is implemented as a bridge. Clients directly connect to a bridge before traffic is forwarded into the Tor network in the typical way. This limits defenses in WFDefProxy to protecting against network adversaries between client and bridge and not against adversaries in the Tor network or in control of the bridge (i.e., the bridge is trusted). Defenses are implemented at a high level in the Go programming language, providing richer features than Maybenot machines, padding machines in the Tor Circuit Padding Framework, and the regularization algorithm in QCSD.

%% file: src/conclusion.tex
We presented version 2 of Maybenot, a framework for traffic analysis defenses heavily inspired by the Tor Circuit Padding Framework~\cite{circpad,circpaddoc}. Defenses are implemented as probabilistic finite-state machines, and the framework provides a common interface for integrating them into protocols such as Tor~\cite{tor}, Wireguard~\cite{wg}, and QUIC~\cite{quic}. Maybenot is implemented as a Rust library (crate) with the goal of being easy to integrate into new and existing protocols. To assist in the development of defenses, we provide a simulator that can be used to simulate how provided network traces may change if given machines were running at the client and/or server.

Our goal with Maybenot is to contribute towards the widespread real-world use of traffic analysis defenses. We hope that Maybenot will be useful for researchers, protocol developers, and defenders alike. With the monumental progress being made in AI and machine learning, we believe that traffic analysis defenses will become increasingly important. Because we are in the middle of this AI revolution, a framework is likely worthwhile in the short to medium term until the dust settles. It took us decades to get to where we are today with making encrypted end-to-end communication the norm. We will probably need a similar amount of time to get to where we want to be with traffic analysis defenses.